\documentclass[aps,preprintnumbers,amsmath,twocolumn,groupedaddress,superscriptaddress,notitlepage,nofootinbib,prl]{revtex4-1}
\usepackage[utf8]{inputenc}
\usepackage{indentfirst}
\usepackage{amssymb}
\usepackage{amsmath,latexsym}
\usepackage{graphicx}
\usepackage{float}
\usepackage[margin=1.0in]{geometry}
\usepackage[mathscr]{euscript}
\usepackage{color}
\usepackage[dvipsnames]{xcolor}
\usepackage{soul}
\usepackage{slashed}
\usepackage{feynmf}
\usepackage{subcaption}
\usepackage{braket}
\usepackage{ulem}
\usepackage{comment}

\usepackage{dsfont}
\newcommand{\bbid}{\mathds{1}}
\newcommand{\beq}{\begin{equation}}
\newcommand{\eeq}{\end{equation}}
\newcommand{\bea}{\begin{eqnarray}}
\newcommand{\eea}{\end{eqnarray}}

\long\def\beqs#1\eeqs{\beq\begin{split} #1 \end{split}\eeq}

\newcommand{\bo}{\mathbf}

\definecolor{MyRed}{RGB}{153,0,13}

\usepackage[colorlinks=true,backref=true,linktocpage=true,citecolor=MyRed,urlcolor=MyRed,linkcolor=MyRed,pdfpagemode=UseOutlines]{hyperref}

\hypersetup{%
  bookmarksnumbered=true,
  pdftitle = {},
  pdfsubject = {},
  pdfauthor = {},
  pdfkeywords = {}
}

\usepackage{microtype}

\usepackage{pgfplotstable,array,siunitx}
\usepackage{multirow}
\usepackage{booktabs}

\usepackage{tikz}

\preprint{INT-PUB-20-030}

\newcommand{\fig}[1]{Fig.~\ref{#1}}

\begin{document}
\title{Universality of a truncated sigma-model}

\author{Andrei Alexandru}
\email{aalexan@gwu.edu}
\affiliation{Department of Physics,
The George Washington University, Washington, DC  20052}
\affiliation{Department of Physics,
University of Maryland, College Park, MD 20742}

\author{Paulo F. Bedaque}
\email{bedaque@umd.edu}
\affiliation{Department of Physics,
University of Maryland, College Park, MD 20742}

\author{Andrea Carosso}
\email{acarosso@gwu.edu}
\affiliation{Department of Physics,
The George Washington University, Washington, DC  20052}

\author{Andy Sheng}
\email{asheng@umd.edu}
\affiliation{Department of Physics,
University of Maryland, College Park, MD 20742}

\preprint{}

\date{August 19, 2021}

\begin{abstract}

Bosonic quantum field theories, even when regularized using a finite lattice, possess an infinite dimensional Hilbert space and, therefore, cannot be simulated in quantum computers with a finite number of qubits. A truncation of the Hilbert space is then needed and the physical results are obtained 
after a double limit: one to remove the truncation and 
another to remove the regulator (the continuum limit). A simpler alternative is to find a model with a finite dimensional Hilbert space  belonging to the same universality class as the continuum model (a ``qubitization"), so only the space continuum limit is required. A qubitization of the $1+1$ dimensional asymptotically free $O(3)$ nonlinear $\sigma$-model based on ideas of non-commutative geometry was previously proposed~\cite{Alexandru:2019ozf} and, in this paper, we provide evidence that it reproduces the physics of the $\sigma$-model both in the infrared and the ultraviolet regimes.

\end{abstract}

\maketitle

\section{Introduction}
A natural way of simulating quantum systems with quantum computers is to map the physical degrees of freedom into a quantum register and evolve its qubits  in time according to the actual Hamiltonian $H$ of the physical system. Assuming that the time steps do not require an exponentially large number of quantum gates, as it is the case for local Hamiltonians, this procedure is exponentially faster than the classical computation of $e^{-iHt}$ which would require the diagonalization of the Hamiltonian matrix that is exponentially large on the number of physical degrees of freedom.

The encoding of the physical degrees of freedom in qubits is subtle in the case of systems described by quantum field theories.
The total Hilbert space of a continuum QFT is infinite dimensional, in general, 
since it is the tensor product of the Hilbert space at each point and
there are an infinite number of points in space.
In addition, in bosonic theories, the local Hilbert space corresponding to a single point is itself infinite dimensional. Quantum computers, however, have a finite number of qubits and are described by a finite dimensional Hilbert space. It is clear that some approximation is required to simulate field theories in quantum computers. The first infinity, caused by the infinite number of points in space, can be regularized by discretizing space onto a lattice with spacing $a$. For bosonic theories, one must also regularize the \textit{field space} of the theory, in order to render the 1-site Hilbert spaces finite. Thus, physical results are expected to be obtained only after taking the double limit of $a\to 0$ and removal of the field space truncation. This double limit is not only cumbersome but is, in many cases, not even possible as the truncation cannot be made arbitrarily fine (see, e.g. \cite{Alexandru:2019nsa}).

The approximation of space by a lattice is well understood from lattice field theory studies and is based upon the concepts of renormalization and universality. It is found that there is great latitude in choosing the discretized theory, as different discretizations sharing the same symmetries lead, in the continuum limit, to the same theory.
This observation suggests that the double limit above may not be necessary. As long as the field space truncation does not alter the universality class of the model, the full model is recovered as $a\rightarrow 0$ without the necessity of eliminating the field space truncation\footnote{If the qubit regularization is thought of as a spin chain, $a\to 0$ constitutes the \textit{quantum} critical point of the system.}. In a situation where qubits are expensive, this is a great boon.

An example of this idea was offered in~\cite{Alexandru:2019ozf}. The $O(3)$ nonlinear $\sigma$-model has a field taking values on the sphere $S^2$. The substitution of the sphere by any set of points~\cite{Patrascioiu:1997ds,Hasenfratz:2000sa}, in a similar vein with~\cite{Alexandru:2019nsa}, leads to a model that is likely different from the $O(3)$ $\sigma$-model~\cite{Caracciolo:2001jd}. This truncation reduces the symmetry and is less useful.
If, however, $S^2$ is substituted by a ``fuzzy sphere" (a construction from non-commutative geometry used previously in different contexts \cite{Madore:1991bw}), the full $O(3)$ symmetry is maintained and it is reasonable to expect the fuzzy model to reproduce the $\sigma$-model in the space continuum limit. However, as stressed in \cite{Singh:2019uwd,Bhattacharya:2020gpm}, in $1+1$ dimensions where the $O(3)\ \sigma$-model is asymptotically free, symmetries alone might not guarantee that a particular qubitization is enough to reproduce the continuum theory in both its infrared and ultraviolet regimes. In fact, cases where the ultraviolet behavior of the $\sigma$-model is and is not reproduced are shown in \cite{Bhattacharya:2020gpm}. In this paper we use matrix product states and finite-size scaling to give evidence that the fuzzy model with \textit{anti}-ferromagnetic coupling indeed reproduces the $O(3)\: \sigma$-model at all energy scales from the infrared, $E \ll m$, to the ultraviolet, $E\gg m$. Here $m$ is the mass gap, the intrinsic scale of the theory.

\section{The Fuzzy $\sigma$-model}

The fuzzy sphere is obtained by replacing the function space on the sphere $S^2$ by a finite-dimensional vector space of matrices. If $\bo n = (n_i)$ is a unit vector on $S^2$, one maps $n_i \mapsto J_i$ and $\partial_{n_i} \mapsto [J_i, \bullet]$, where $J_i, \; i=1,2,3$, are the generators of a spin-$j$ $SU(2)$ representation, for any desired $j$ \cite{Madore:1991bw}. Functions on the sphere $\psi(\bo n)$ are then mapped via their Taylor expansion in $n_i$ to $(2j+1)\times(2j+1)$ matrices $\Psi$. For any fixed $j$, the Taylor expansion for $\Psi$ terminates at finite order, yielding a finite $(2j+1)^2$-dimensional Hilbert space. In the case $j=1/2$, the $J_k$ are proportional to the Pauli matrices $\sigma_k$. The 1-site Hilbert space is then 4-dimensional, with states of the form
\beq
\Psi = \psi_0\bbid + \sum_{k=1}^{3}\psi_k J_k.
\eeq
In the orthonormal basis $\{|a\rangle, a=0,1,2,3\}=\{i/\sqrt{2} \bbid,\sqrt{3/2} J_k\}$,\footnote{The inner product of two fuzzy states is $\langle \psi| \phi \rangle \equiv \mathrm{tr}[\psi^\dag \phi]$.} the Hamiltonian takes the form
\beq \label{Hamiltonian}
H = \eta\Big[g^2\sum_{x}h_{0}(x) \pm \frac{3}{4g^2} \sum_{x,k} j_{k}(x) j_{k}(x+1)\Big],
\eeq
with  $h_0 = \rm{diag} (0,1,1,1)$ and
\beq
j_1 = \frac{\bbid\otimes \sigma_2}{\sqrt{3}}, \quad j_2 = \frac{\sigma_2 \otimes \sigma_3}{\sqrt{3}}, \quad j_3 = \frac{\sigma_2 \otimes \sigma_1}{\sqrt{3}},
\eeq
where $(j_k)_{ab} = \langle a | J_k | b\rangle$ are $4\times 4$ matrices. We take $g^2>0$; the $\pm$ sign in $H$ refers to the anti-ferromagnetic and ferromagnetic cases, respectively. The Hamiltonian is invariant under rotations.

The parameter $\eta$ linearly  scales  the energies in the spectrum of $H$, without affecting the eigenstates $\ket{\Psi_k}$ (and the spatial correlation length of the theory). Its value can be set in such a way that the energy gap $a\Delta = aE_1 - aE_0$ equals  the inverse spatial correlation length $am = \xi^{-1}$ of the system -- a condition required by a relativistic theory~\cite{Shigemitsu:1981}. The continuum limit is obtained by sending $g^2\rightarrow 0$ along lines in the $(g^2,\eta)$ plane such that the energy gap coincides with $m$.

To determine $\eta$ for a given $g^2$ we compute  the gap $a\Delta=aE_1-aE_0$ using $\eta=1$. The mass $am=1/\xi$ is determined by extracting the correlation length from the dependence on $x$ of the spatial correlator,
\beq \label{corr}
C(x) = \langle \Psi_0 | J_3(0)J_3(x) | \Psi_0 \rangle\,,
\eeq
where $\ket{\Psi_0}$ is the ground state (which is independent of $\eta$). We then set $\eta=am/a\Delta$.

\section{Matrix product states}


We investigate the system numerically using a variational matrix product state (MPS) algorithm to compute the low-lying eigenstates and eigenvalues of the Hamiltonian~\cite{White:1993}. That is, we assume the following ansatz for the states of an $N$-site system
\beq
|\Psi_A \rangle = \sum_{a_1, \dots, a_N} \mathrm{tr} \big( A^{a_1}_1 \cdots A^{a_N}_N \big) | a_1, \dots, a_N \rangle,
\eeq
where $a_n = 0,1,2,3$, and the $A^{a_n}_n$ are $D\times D$ matrices, and then iteratively minimize the expectation value of $H$ with respect to the $A^{a_n}_n$. This procedure yields an approximation for the ground state $|\Psi_0\rangle$ and its energy $aE_0$. Excited states $|\Psi_k\rangle$ and energies $aE_k$ are obtained via a similar algorithm with the additional constraints $\langle \Psi_j|\Psi_k\rangle_{j<k} = 0$. 
This approximation becomes exact as $D\to 4^N$, but it converges quickly, at least for gapped systems.

Although the Hamiltonian we consider here uses periodic boundary conditions, algorithms for periodic MPS are less efficient, having a computational cost of order $O(D^5)$, while algorithms for open-boundary MPS ansatz, where $A_1$ and $A_N$ are $1\times D$ matrices, have a cost of order $O(D^3)$~\cite{Pippan:2010}.
We stress that only the \textit{states} in the open-boundary case are not periodic; the Hamiltonian remains periodic.
We have checked in several cases that the low-lying energies from the open-boundary ansatz are consistent with results from the periodic ansatz up to a relative error of order $10^{-4}$--$10^{-6}$,  with clear convergence toward the open-boundary value as $D$ is increased.
We have also checked for a few values of $g^2$ that the spatial correlation lengths $\xi$ obtained from both open-boundary and periodic MPS ansatz agree to a few parts per thousand, by directly computing the correlators in the MPS ground states in both cases. 
Although the open-boundary MPS ansatz breaks translation invariance, the correlator agrees well with the periodic one away from the boundaries.

\section{Finite-Volume Mass Gaps}

In the infinite-volume limit, the rescaled energy gap is the mass of the particles in the model. In a finite-volume box of size $L$ with periodic boundary conditions, $\Delta$ is corrected by the interactions between the particle and its periodic images. The volume dependence of $\Delta$ can be used as a convenient probe of the theory at an energy scale set by $1/L$ since it depends on the phase shifts of two-particle scattering at energies $\sim 1/L$, as derived by Lüscher~\cite{Luscher:1986pf}.
Here, we compare the continuum limit results for our model with these predictions, first by focusing on one kinematic point and then more broadly by comparing the step-scaling curve of the fuzzy model with that of the $\sigma$-model. 

We first discuss the energy gap in the infrared limit ($mL \gg 1$), since this
allows us to compute the parameter $\eta$ directly from the energy gaps,
which is computationally cheaper than computing the correlator in \eqref{corr}. 
The $S$-matrix elements are known for the (1+1)-dimensional $\sigma$-model~\cite{1978NuPhB.133..525Z} and so the explicit form of the gap volume dependence is known~\cite{Klassen:1990ub},
\beq \label{lusch_gaps}
a\Delta(L) = \frac{am}{\eta}\big(1 + I(mL)\big) + \mathcal{O}(e^{-\sqrt{3}mL}),
\eeq
where
\beq
I(mL) = 4\pi\int_{-\infty}^{\infty} d\theta e^{-mL\cosh\theta} \frac{\cosh\theta}{\theta^2 + \frac{9\pi^2}{4}}.
\eeq

For a given $g^2$, we calculate the gaps $a\Delta(L/a)$ for several 
system sizes $N=L/a$. We then fit the data to the L\"uscher formula \eqref{lusch_gaps} in the range of the formula's validity to extract $am$ and $\eta$. Errors are assigned to the gaps based on the convergence rate of the variational algorithm used to minimize the MPS energy. For $D$ $\sim$ 500--1500, depending on the system size, the relative error for the gaps is estimated to be about one in $10^{4}$. In the $mL\gtrsim 3$ regime we get good fits to the L\"uscher formula.
We repeat this at several different $g^2$ and interpolate using quadratic ansatze to get $am(g^2)$ and $\eta(g^2)$, so that we can easily tune the value of $g^2$ for a required $am$ or $\eta$.

In principle, a determination of $am$ from the correlators and an estimate of the infinite-volume gap $a\Delta$ is sufficient to provide $\eta$. However, because the energy estimates from the MPS algorithm converge more quickly than (the long distance structure of) the states, we choose the Lüscher formula methodology of obtaining $am$ and $\eta$ over that of the correlator. At several $g^2$, we have checked that $am$ obtained from the Lüscher formula agrees with that computed from fitting the spatial correlator, to several parts per thousand.

\begin{figure}[t!]
\centering
\begin{minipage}{.5\textwidth}
\includegraphics[width=1.0\textwidth]{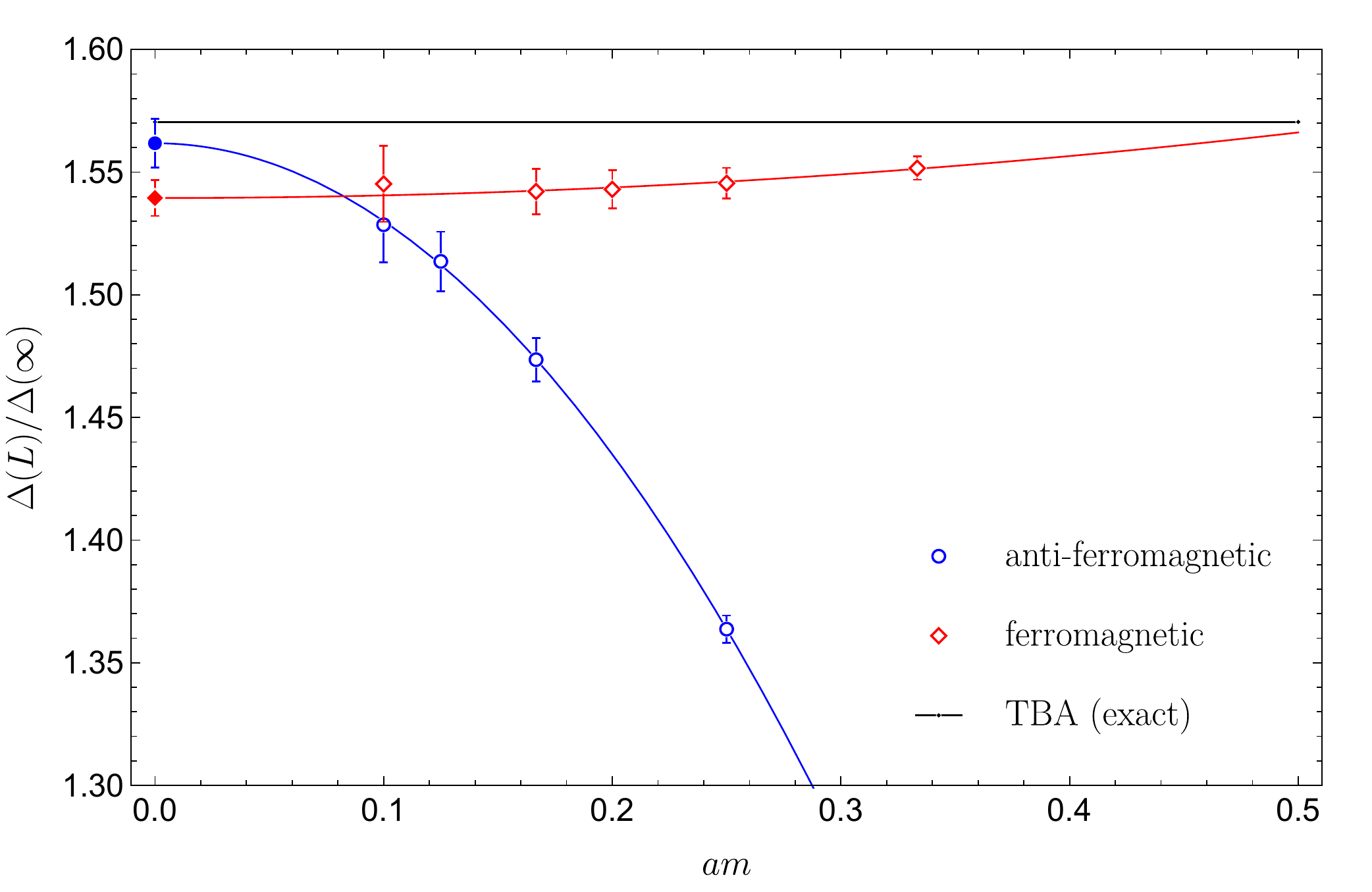}
\caption{\small{Continuum limit of the finite-volume mass gap at fixed $mL = 1$ for both fuzzy models. The curve fits assume a discretization error of order $a^2$. Note that $m$ is the fixed, physical mass, and here $ma=1/N$. \label{fig:finite_volume_gap}}}
\end{minipage}\hfill
\end{figure}

\begin{figure*}[t!]
\includegraphics[width=0.49\textwidth]{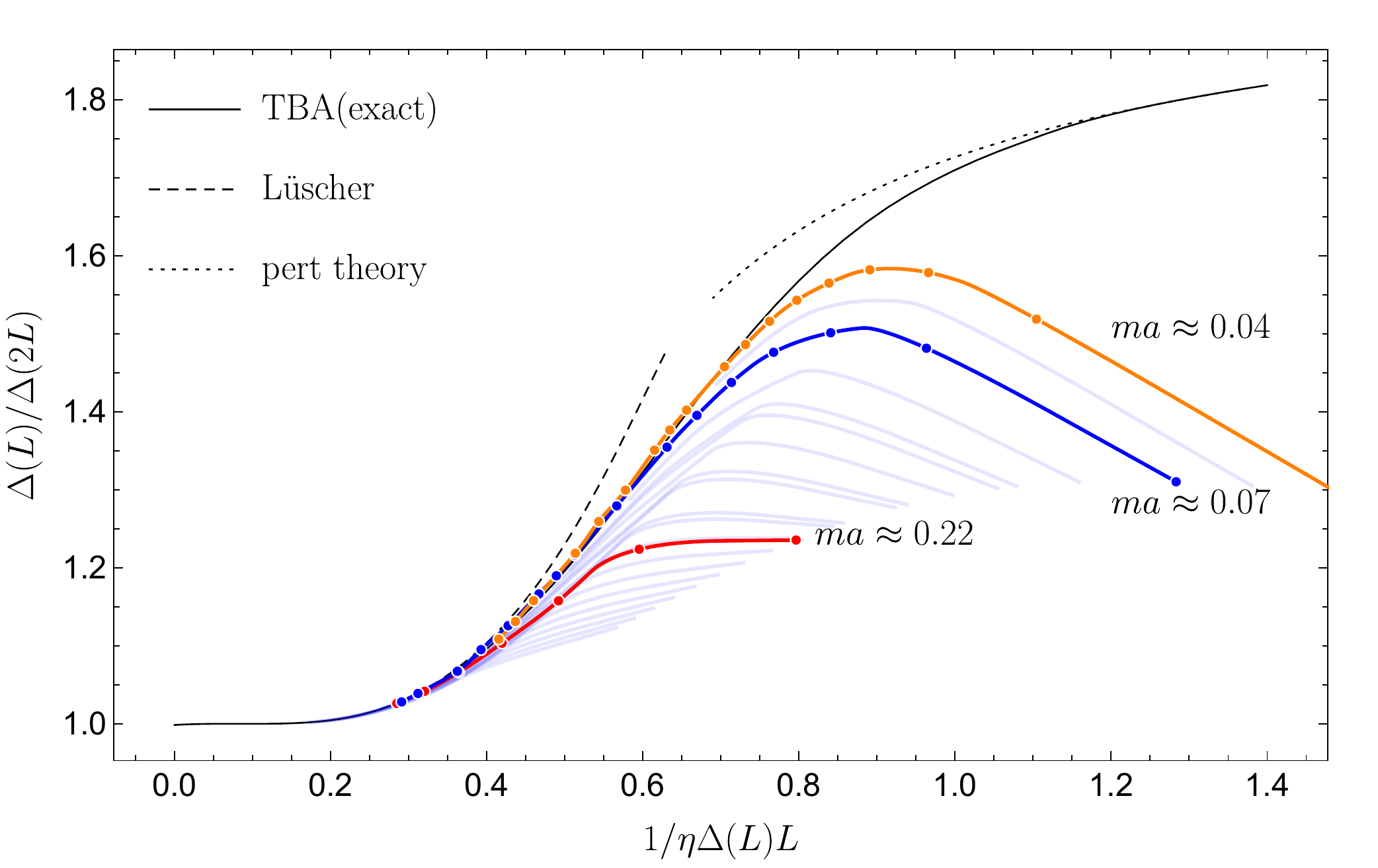}
\includegraphics[width=0.49\textwidth]{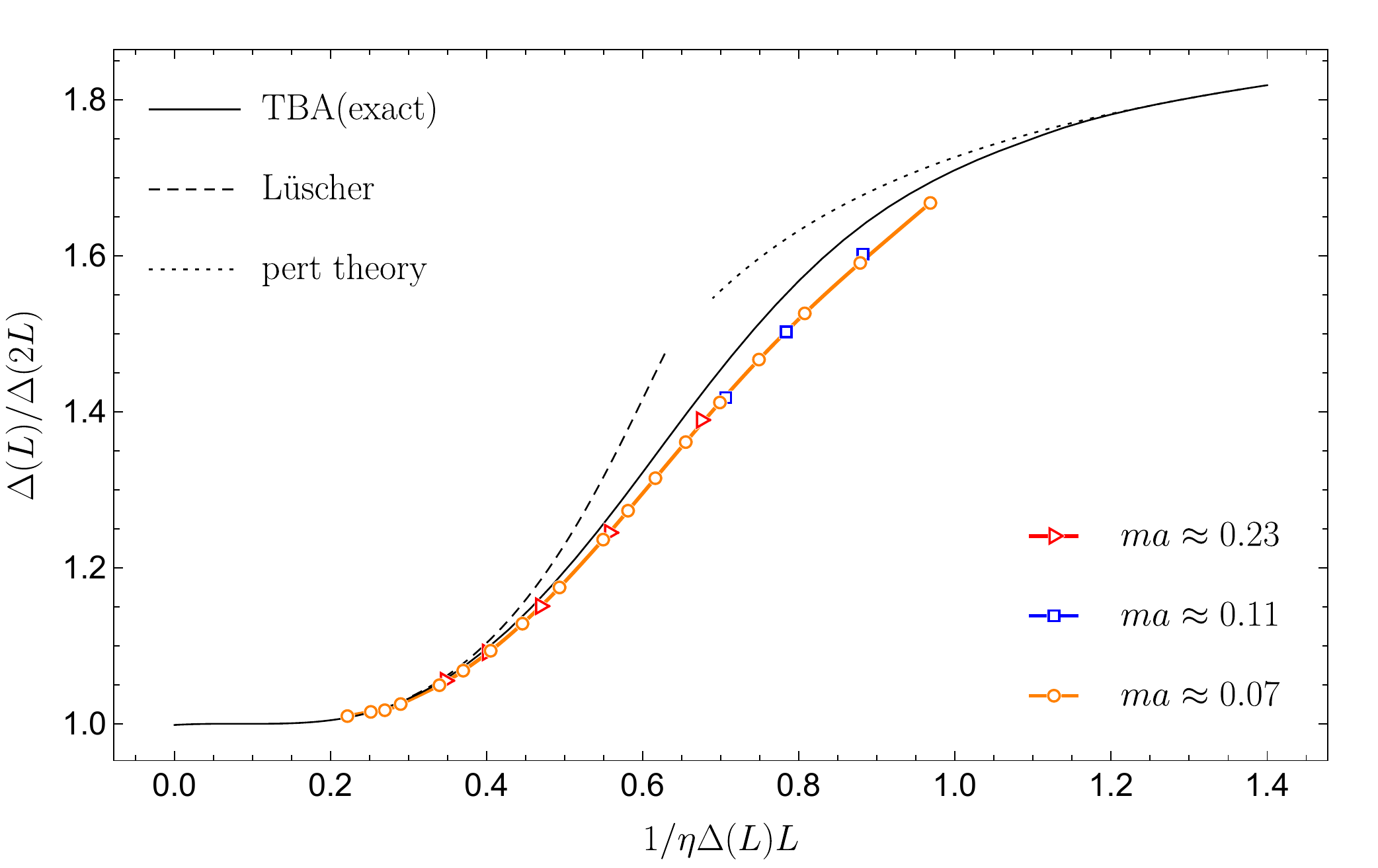}
\caption{\small{Step-scaling curves at constant $g^2$ for the anti-ferromagnetic (left) and ferromagnetic fuzzy models (right), compared with scaling curves obtained from the L\"uscher formula, TBA, and 3-loop perturbation theory. We include interpolations between the data points to guide the eye. \label{fig:caracciolo}}}
\end{figure*}

As a preliminary check, we compare the finite-volume energy of the fuzzy model with exact results for the $O(3)$ $\sigma$-model derived using the thermodynamic Bethe ansatz (TBA)~\cite{Hasenfratz:1990zz,Balog:2003yr}.
To get the continuum results, we carry out simulations for
a set of increasing values of $L=Na$, while keeping $mL$ fixed. For
every $N$ we tune $g^2$ to get the desired $am=mL/N$, and calculate
the value of the gap $a\Delta(L)$. Since the system size is fixed,
the continuum limit is achieved for $N\to\infty$. The discretization
errors are expected to vanish as $a^2$ which agrees with the convergence
rate for our results.

For kinematic points in the infrared, $mL=4$ and 6, we checked that both the ferromagnetic and anti-ferromagnetic models agree with predictions within the error-bars. This is not surprising since in this regime, the exact results agree very well with \eqref{lusch_gaps}, with which we performed our fits. However, \eqref{lusch_gaps} loses its validity as $mL$ becomes small.
In \fig{fig:finite_volume_gap} we extrapolate the fuzzy mass gaps to the continuum at $mL = 1$ and compare them with the TBA prediction. The extrapolations suggest that the anti-ferromagnetic model is consistent with the TBA results whereas for the ferromagnetic model, there is some tension. This indicates that while the ferromagnetic model agrees in the infrared with the $\sigma$-model it likely does not belong to 
the same universality class.



A more stringent check involving a broader kinematic range is a comparison of the step-scaling curves. Step-scaling measures the response of a suitably defined
finite-volume correlation length to a doubling of the the system size. Here we define
$\xi(g^2, L) = 1/\eta(g^2) \Delta(g^2, L)$ and the step-scaling curve is
$\xi(g^2, 2L)/\xi(g^2, L) = \Delta(L)/\Delta(2L)$ as a function of $\xi(L)/L = 1/\eta \Delta(L) L$. In the continuum limit this should be the same curve for all models within the same universality class~\cite{Caracciolo:1994,Caracciolo:1995}. In the infrared limit $mL\rightarrow \infty$, $\Delta(L)/\Delta(2L) \rightarrow 1$ as the energy of the one-particle state becomes independent of the volume (the dependence is exponentially small on the volume). In the opposite limit $mL\rightarrow 0$, $\Delta(L)/\Delta(2L) \rightarrow 2$ since $\Delta(L) \propto 1/L$~\cite{Shin:1996gi}. In~\fig{fig:caracciolo}, we plot the zero temperature scaling curve of the $O(3)$ $\sigma$-model determined from the TBA calculation of~\cite{Balog:2003yr}, along with the infrared curve determined by the Lüscher formula \eqref{lusch_gaps}, and the ultraviolet curve from perturbation theory in the asymptotically free regime~\cite{Shin:1996gi}.\\\\
\indent We computed the scaling curves for both ferromagnetic and anti-ferromagnetic fuzzy models and the results are shown in \fig{fig:caracciolo}. For clarity, curves are interpolated between points with constant $g^2$. In the infrared limit, all curves converge toward the TBA and Lüscher curves. In the opposite limit, the two models exhibit distinct behavior. In the anti-ferromagnetic case, the curves hug the TBA line and eventually peel off at small $L$ where discretization effects become important. As the continuum limit is approached, $g^2 \to 0$, the $g^2$-constant curves continue to follow the scaling curve deeper into the ultraviolet. In the ferromagnetic case, however, there is a qualitatively different behavior whereby all $g^2$-constant curves overlap, while deviating  notably from the TBA curve as they leave the infrared region. Thus the scaling curve of the anti-ferromagnetic fuzzy model agrees with the $\sigma$-model over a wide range of scales, while the ferromagnetic model seems to behave differently for higher energies.



\section{Conclusions}

The (1+1)-dimensional $O(3)\ \sigma$-model is asymptotically free and describes a triplet of particles with mass $m$. 
We showed that the finite-volume energies computed using the anti-ferromagnetic fuzzy model agree to high precision with the exact predictions for the $\sigma$-model, both in the infrared region where the Lüscher formula approximates finite-volume corrections well and  in the ultraviolet regime where the formula breaks down. Since the finite-volume corrections are controlled by the dynamics of the $\sigma$-model, this agreement provides strong evidence that the anti-ferromagnetic fuzzy model shares the same universality class as the $O(3)$ $\sigma$-model in 1+1 dimensions, and therefore constitutes a qubit regularization of the bosonic field theory that may be used on quantum computers. Note that agreement in the far infrared region (distances much larger than $1/m$) is already accomplished by the spin-1 anti-ferromagnetic Heisenberg chain \cite{Haldane:1982rj}. It is the agreement at smaller distance scales, where asymptotic freedom kicks in, that is the noteworthy property of the fuzzy model (and one of the models in \cite{Bhattacharya:2020gpm}).

For the ferromagnetic fuzzy model we find that the results are in tension with the $\sigma$-model. However, we note that the ferromagnetic case has a mass gap and the scaling curve seems to converge rather quickly to a different continuum limit than the $\sigma$-model outside the deep infrared; this is perhaps unexpected since the analogous ferromagnetic spin-1 chain is gapless in the continuum. The identification of the universality class of the ferromagnetic fuzzy model is therefore left for a future study.

\section{Acknowledgments}

We thank J\`anos Balog for providing us with the TBA data for the step-scaling curve of the $O(3)$ model. A.A. and A.C. are supported by U.S. DOE Grant No. DE-FG02-95ER40907. P.B and A.S. are supported in part by the US DoE under contract No. DE-FG02-93ER-40762. P.B. and A.A. were  also supported by U.S. DOE Grant No. DE-SC0021143

\bibliographystyle{apsrev4-1}
\bibliography{fuzzy-mps.bib}

\end{document}